\documentclass[12pt]{iopart}
\usepackage[dvips]{graphicx}
\usepackage[utf8]{inputenc}
\usepackage[usenames,dvipsnames]{xcolor}
\usepackage[colorlinks=true, linkcolor=RoyalBlue, citecolor=RoyalBlue]{hyperref}

\newcommand{\subref}[2]{\ref{#1}{\color{RoyalBlue}({#2})}}
\newcommand{\textfrac}[2]{{#1}/{#2}}

\begin{document}
\title[Magnetotransport in disordered 2d TIs: signatures of charge puddles]{Magnetotransport in disordered two-dimensional topological insulators: signatures of charge puddles}

\author{Sven Essert$^1$ and Klaus Richter$^1$}
\address{$^1$ Institut f\"ur Theoretische Physik, Universit\"at Regensburg, D-93040 Regensburg, Germany}

\ead{Sven.Essert@ur.de}

\begin{abstract}
In this numerical study we investigate the influence and interplay of disorder, spin-orbit coupling and magnetic field on the edge-transport
in HgTe/CdTe quantum wells in the framework of coherent elastic scattering. We show that the edge states
remain unaffected by the combined effect of moderate disorder and a weak magnetic field at realistic
spin-orbit coupling strengths. Agreement
with the experimentally observed linear magnetic field dependence for the conductance of long samples is obtained
when considering the existence of charge puddles.
\end{abstract}

\maketitle

\section{Introduction}
Both from a fundamental and from an applications viewpoint, two-dimensional (2d) topological insulators (TI) are very promising because of their quasi-one dimensional
(1d) edge states, which are (unlike many other 1d quantum systems) protected against localization due to
elastic disorder scattering.  This protection is guaranteed by time-reversal symmetry. Hence, one question that quite naturally
arises is how fast the edge channels exhibit backscattering and eventually localize under the application of an external magnetic field.

The first published experiments on HgTe/CdTe quantum wells \cite{Konig2007,Konig2008} showed an almost linear dependence of the edge conductance
on the magnetic field and suggested a strong suppression of the edge transport already for low field strengths which
was reproduced by other groups \cite{Gusev2014a}. 
These measurements were, however, performed with samples much larger than the coherence length which did not show a quantized conductance at zero field.
For smaller samples in which the transport is assumed to be fully coherent, the conductance
remains unaffected by the magnetic field up to high field strengths \cite{Buhmanncomm}. 
Correspondingly, in the first experiments of InAs/GaSb based TI the transport in small systems turned out to be robust
while long non-quantized samples were more fragile with respect to the magnetic field \cite{Knez2011}. However, recent experiments in
this material system show that one can also produce long samples which are robust up to at least 1 Tesla \cite{Knez2014}.

On the theory side, there are already a few publications dealing with different aspects of the edge-transport properties of 2d-TIs (mainly focusing on HgTe/CdTe) under
the influence of a perpendicular magnetic field using various methods \cite{Maciejko2010,Chen2012a,Xue2012,Delplace2012,Pikulin2014,Essert2014}. We will not discuss them all 
in detail but we would like to mention a calculation by Maciejko \etal \cite{Maciejko2010} which considers the combined effect of elastic disorder scattering 
and applied magnetic field.
Within this model they find a strong and almost linear magnetic field dependence of the conductivity, very similar to the findings for the
experiments on long HgTe/CdTe samples but different from results on the short HgTe/CdTe samples in which the quantized conductance is observed.
This is surprising as this theory assumes a fully coherent transport picture, something that one would rather assume to be fulfilled
for the short but not for the longer samples. In addition, there are later theory calculations which use a comparable calculation setup \cite{Chen2012a,Pikulin2014} 
but show a much weaker influence of the magnetic field.

Here we present a systematic study, including elastic disorder scattering, spin-orbit interaction and magnetic field. Thereby, we try to resolve this discrepancy, 
and find out whether it is possible to describe both scenarios (long and short samples). 

Note that we are restricting our scope to the influence of a moderate perpendicular magnetic field. We will, e.g., not consider band structure changes
which are expected at much larger fields leading to a phase transition of the TI to a quantum hall phase \cite{Scharf2012}.

In the first part of this manuscript, we have a close look at the influence of elastic backscattering and disorder on the edge transport in
HgTe/CdTe ribbons. We find that the effect of the magnetic field is very weak but it becomes stronger at disorder strengths high enough
to make the bulk of the material conductive as it is undergoing a phase transition to a trivial insulator. We show that this disorder-induced
phase transition may be used to experimentally control the edge transport by creating samples with inhomogeneous disorder. In the last part, we reexamine the effect of local 
metallic puddles which are not due to disorder
but due to locally gated regions which may be naturally present owing to trapped charge impurities. The transport through such puddles does
indeed show the strong linear dependence that was found by \cite{Konig2008,Gusev2014a}, supporting the fact that this magnetoconductance
signature may be a hint to the presence of charge puddles in these materials.

\section{Theory}
We use the Bernevig-Hughes-Zhang (BHZ) Hamiltonian \cite{Bernevig2006} to model the electronic structure of HgTe/CdTe quantum wells,
%
%
\begin{equation}
H=\left(\begin{array}{cc}
h(\mathbf{k}) & \begin{array}{cc}
0 & -\Delta\\
\Delta & 0\end{array}\\
\begin{array}{cc}
0 & \Delta\\
-\Delta & 0\end{array} & h^{*}(-\mathbf{k})\end{array}\right),
\label{eq:bhzhamilton}
\end{equation}
%
%
with spin-subblock Hamiltonians
%
%
\begin{equation}
h(\mathbf{k})=\left(\begin{array}{cc}
M-(B+D)\mathbf{k}^{2}+V & Ak_{+}\\
Ak_{-} & -M+(B-D)\mathbf{k}^{2}+V\end{array}\right),
\end{equation}
%
%
where $k_{\pm}=k_{x}\pm \rmi k_{y}$ and $\mathbf{k}^{2}=k_{x}^{2}+k_{y}^{2}$. The material parameters $A$, $B$, $D$, $M$ and $\Delta$ are
taken from $k\cdot p$-calculations \cite{Konig2008}. Here, $\Delta=1.6\,\mathrm{meV}$ is accounting for the spin-orbit coupling resulting
from bulk inversion asymmetry. $V$ includes the effects of an electrostatic potential, which will be applied in the second part
of this publication.
We do not include Rashba spin orbit coupling, the size of which is strongly depending on the detailed layer structure of the quantum well.
However, we did some test calculations that show that the effects presented in this manuscript are qualitatively unchanged under the inclusion
of an additional Rashba coupling term. For similar reasons, we also do not include Zeeman terms as we found that they are of minor importance
for the scenario that we discuss, i.e., transport in a moderate perpendicular magnetic field.

%
%
\begin{figure}
\centering
\includegraphics[width=.7\textwidth]{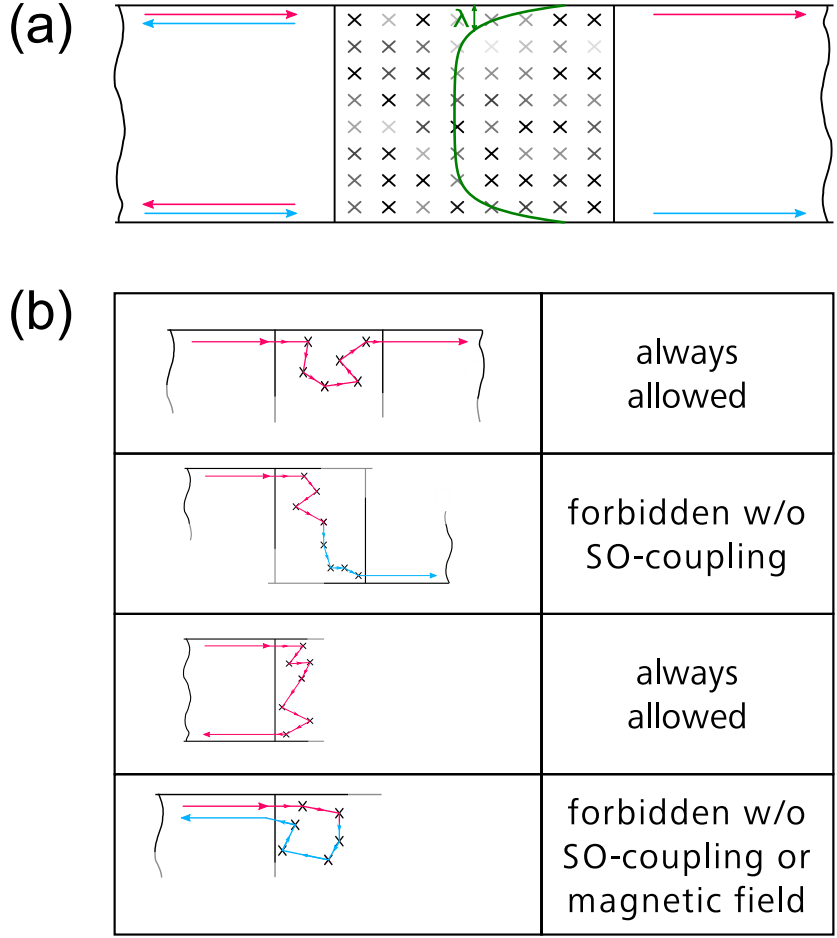}
\caption{ (a) Scheme of the ribbon geometry used for transport calculations (different spin orientations are marked in blue and red). The Fermi energy is chosen such that there 
are only edge states present in the leads. The meaning of $\lambda$, cf. \eref{eq:lambda}, is shown on an exemplary transverse local density of states profile (green).
(b) The table lists all the possible scattering paths in the above geometry (changes of color along a path indicate spin flip due to spin-orbit (SO) coupling). In addition, it shows how some of these paths are symmetry
forbidden in the absence of spin-orbit coupling or magnetic field. Scattering paths through
the bulk (to the other edge, e.g. in the third panel) will be strongly suppressed due to the bulk
being insulating as long as the disorder strength is small enough.
}
\label{geostrip}
\end{figure}
%
%

For the numerical transport calculations this Hamiltonian is discretized on a regular grid with lattice constant $a = 5\,\mathrm{nm}$. The
disorder is taken to be of Anderson type, meaning that it will be on-site disorder drawn from a box-distribution $V_\mathrm{dis}=\left[-W,W\right]$
which enters the Hamiltonian on the same footing as the electrostatic potential. It is important to note at this point that the ``effective strength''
of the disorder, e.g., the influence of the disorder on the mean free path, is not given by the amplitude $W$ alone but includes
the lattice discretization which is the inherent correlation length of the on-site disorder. The effective strength is then roughly given
by $a^2 W$. This is important when comparing disorder strengths across simulations which use a different lattice spacing.
Also this means that one should not attempt to compare the mere amplitude of the disorder strength with other energies, like the band
gap of the material as this amplitude will vary with the (arbitrarily chosen) discretization. It would therefore be useful not to just
plot the amplitude but instead always use the combination $a^2 W$ as an indicator of the effective disorder strength. However,
we did not do this in this publication to simplify the comparison with phase diagrams calculated in other publications which also
use the same lattice constant as we do ($a=5\,\mathrm{nm}$).

The magnetic field is implemented as a Peierls phase for the hopping matrix elements. In the first part of this publication we consider
a scattering geometry as shown in figure~\subref{geostrip}{a}, i.e., a simple ribbon geometry with clean leads and a central scattering region. As discussed
in the introduction, we are mainly interested in the scattering properties of the edge states under the influence of an external magnetic field.
We therefore choose the Fermi energy of the system in the gap by setting $E_\mathrm{F} = 0\,\mathrm{meV}$ such that only edge states contribute to 
transport in the clean system.

The possible scattering paths in this geometry are listed in figure~\subref{geostrip}{b} together with
their dependence on the presence of spin-orbit coupling and magnetic field. However, the processes that connect the two edges (like the spin-conserving
edge-to-edge reflection which is always allowed by symmetry) will be strongly suppressed as long as the disorder is small enough to
keep up the insulating behavior of the bulk. In this regime (insulating bulk), the on-edge spin-flip backscattering (bottom panel in figure~\subref{geostrip}{b}) is the only
non-trivial scattering pathway but it requires both spin-orbit coupling and a magnetic field. It is for this reason that the
edge state transport is expected to be exceptionally ballistic.

For the above geometry, the low temperature transport coefficients and the local density of states, $d(x,y)$, are calculated with a recursive Green's function 
algorithm \cite{Wimmer2009}. From $d(x,y)$ we extract a quantity which we call the spread of the local density of states, 
\begin{equation}
 \lambda=\frac{\int d\left(x,y\right)\left[y\,\theta\left(\textfrac{L_{y}}{2}-y\right)+\left(L_{y}-y\right)\theta\left(y-\textfrac{L_{y}}{2}\right)\right]\rmd x \rmd y}{\int d\left(x,y\right)\rmd x \rmd y},\label{eq:lambda}
\end{equation}
with $\theta(y)$ being the Heaviside step function. It is a measure
of the distance of the regions with a high local density of states from the sample edge.

\section{Results}
\subsection{Magnetotransport of disordered ribbons}
%
%
\begin{figure}
\centering
\includegraphics[width=0.7\textwidth]{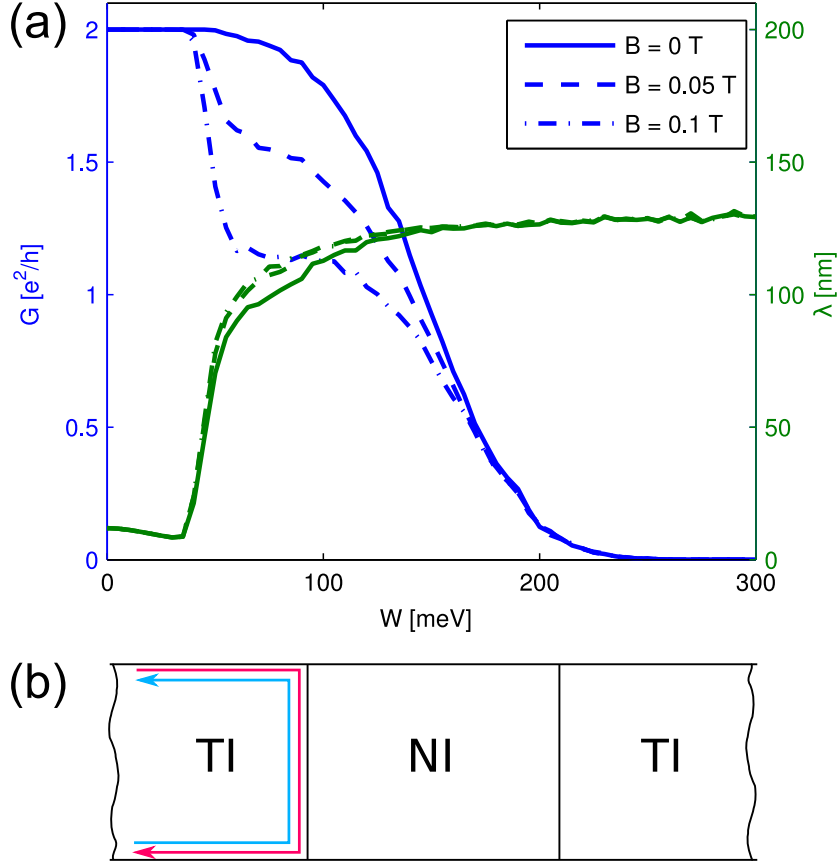}
\caption{ (a) Disorder averaged results of the transmission (blue) and the spread of the local density of states (green) as a function of disorder strength $W$, for
a ribbon geometry as shown in figure~\subref{geostrip}{a}, with $L_x=1000\,\mathrm{nm}$ and $L_y=500\,\mathrm{nm}$. The average was taken over 1000 disorder realizations.
(b) Sketch of the scenario for very strong disorder: The scattering region turns into a trivial Anderson insulator and the edge states are topologically
reflected right at the interface to the scattering region.
}
\label{resrib}
\end{figure}
%
%
In figure~\subref{resrib}{a} typical results are shown for the total transmission as a function of disorder strength (blue lines) through such ribbons for different magnetic fields (shown as
full, dashed, and dash-dotted). One notes that there is no
visible effect of the magnetic field for small disorder strengths ($W < 40\,\mathrm{meV}$): Although the applied fields 
are on the order of a few flux quanta penetrating the ribbon, the transmission stays at the quantized value of $2 e^2/h$.
At around $W \approx 40\,\mathrm{meV}$ the applied magnetic field reduces the conductance. However, from this disorder strength onwards, also the 
conductance at zero field is below the quantized value. This implies that there are also edge-to-edge processes contributing, as pure on-edge-backscattering
is symmetry-forbidden at zero magnetic field, cf. figure~\subref{geostrip}{b}. I.e., the bulk material in this disorder range is no longer a good insulator and
there is a finite edge-to-edge coupling. This is reflected in the spread of the local density of states, $\lambda$, which
is also depicted in figure~\subref{resrib}{a} (green lines).
It shows that the transport is very closely localized along the edge ($\lambda < 15\,\mathrm{nm}$) for small disorder strengths and
undergoes a sharp transition around $W \approx 40\,\mathrm{meV}$. For strong disorder we see a complete spread of the local density of states
($\lambda \approx \textfrac{L_y}{4}=125\,\mathrm{nm}$). Note that also the influence of the magnetic field decreases in this strong disorder regime.

These phenomena can all be understood by considering the phase diagram for disordered HgTe-quantum wells \cite{Prodan2011,Li2009a}. For the Fermi energy at which the previous calculations where performed ($E_\mathrm{F}=0\,\mathrm{meV}$), it shows that 
one expects the bulk material to undergo a phase transition from a topological insulator to a topologically trivial Anderson insulator
for a disorder strength of around $100\,\mathrm{meV}$. This phase transition is of the localization-delocalization type, for which the localization
length diverges at the phase transition. 
For our finite-sized scattering region this implies that there will be a small intermediate metallic regime between the two
insulating phases, for which the localization length is large compared to the sample size\footnote{Interestingly, the position of the phase transition seems to be unaffected by the additional magnetic field, at least up to the magnetic field values that we considered.}.
The effect of the magnetic field seems to be largest in this intermediate metallic regime. It is actually only in
this regime that the on-edge backscattering, the fourth process in figure~\subref{geostrip}{b}, has non-negligible amplitude. This is understandable
in a semi-classical way when thinking of the scattered electrons in terms of diffusive trajectories. Only if they enclose a
sizable amount of flux the time-reversal symmetry can be effectively broken through dephasing thus allowing on-edge backscattering.
This is not possible if the trajectories are bound close to the edge (as in the case of a truly insulating bulk).
Considering the phase transition, one can also understand the behavior in the strong-disorder limit. As the phase
above the transition will be topologically trivial, we face a situation as depicted in figure~\subref{resrib}{b}, which
we dub \emph{topological reflection}. As the leads are clean and topologically non-trivial, topological edge states form along the boundary to the
scattering region. The dominant process will thus be an edge-to-edge reflection, the third process in figure~\subref{geostrip}{b}, which (given that the bulk is sufficiently insulating) will
again involve only one-dimensional states. This explains why the effect of the magnetic field decreases for larger disorder and the spread of
the local density of states approaches that of an even distribution.

%
%
\begin{figure}[]
\centering
\includegraphics[width=0.7\textwidth]{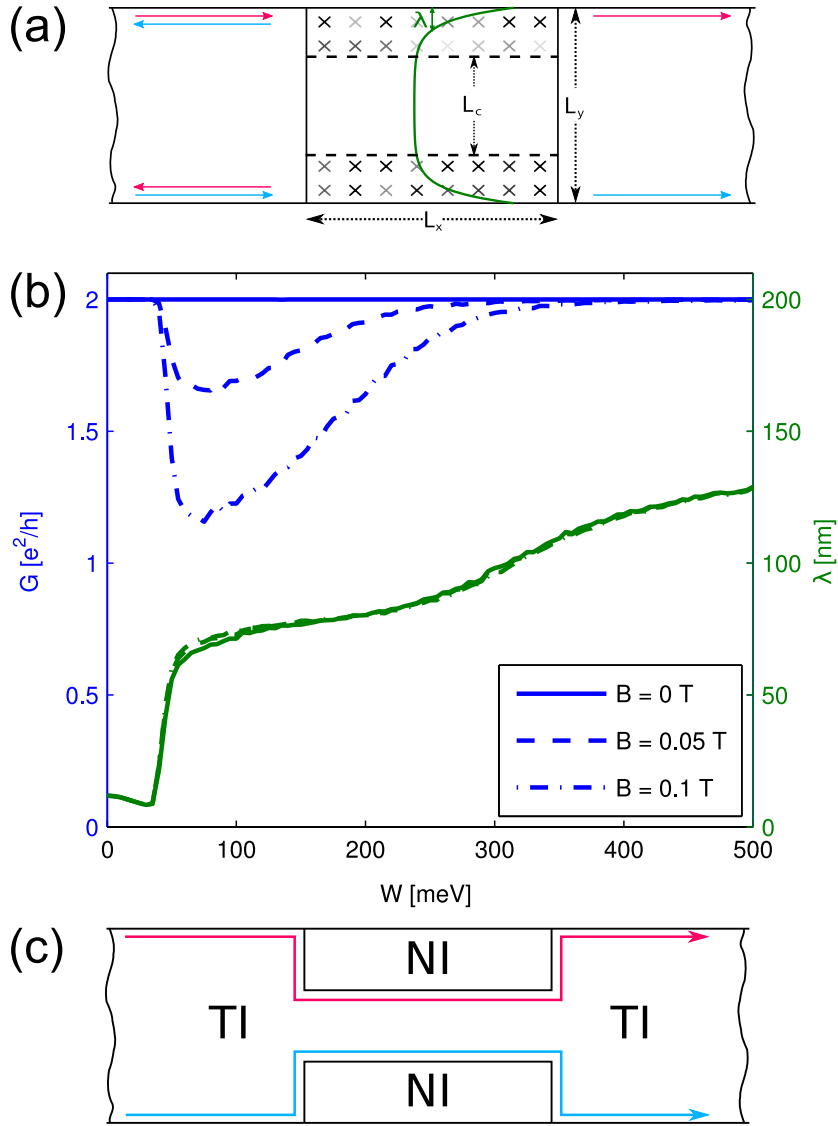}
\caption{ (a) Scheme of the ribbon geometry in which the
disorder is limited to the edges of the sample and a clean gap is left between the two disorder patches on the sides. In this
way the two edges are fully decoupled as there is always a clean insulating region separating them.
(b) Disorder averaged results for the transmission (blue) and the spread of the local density
of states (green), for a gapped geometry as shown in (a),
with $L_x = 1000\,\mathrm{nm}$, $L_y = 500\,\mathrm{nm}$ and $L_c = 200\,\mathrm{nm}$. The average was taken
over 1000 disorder realizations. (c) Sketch of the scenario for
very strong disorder: The disorder regions turn into trivial Anderson insulators and one observes one-dimensional
edge states running around their boundaries.
}
\label{resribcleangap}
\end{figure}
%
%
To fully exclude edge-to-edge backscattering, we choose a calculation setup as shown in figure~\subref{resribcleangap}{a}, i.e., we restrict the disordered
region to the edges of the sample and leave a clean strip in the center of the scattering region. In this way, there will always be a truly
insulating layer that prevents the coupling of the two edges.
Results of these calculations are shown in figure~\subref{resribcleangap}{b}. For low disorder strengths we find similar results as in the previously discussed
setup: There is no effect of the magnetic field on the transport when the disorder strength is low enough that there is pure one-dimensional edge
transport ($W < 40\,\mathrm{meV}$). At around $W\approx 40\,\mathrm{meV}$ the spread of the local density of states increases and reaches an intermediate plateau at around $75\,\mathrm{nm}$ which corresponds
to an equal distribution over the whole disorder region and is a clear sign of the intermediate metallic phase. In this region, the magnetic field
leads to a strong reduction of the transmission. Interestingly, this effect decreases again, for very high disorder strengths ($W > 200\,\mathrm{meV}$). This seems
paradoxical at first glance, as one would expect an increased backscattering at a given magnetic field for higher disorder strengths, but can
also be understood when thinking about the previously mentioned phase diagram and by looking at the spread of the local density of states. In this limit, the value
for $\lambda$ approaches $150\,\mathrm{nm}$ meaning that the transport is now mainly localized at the interface of the clean and the disordered region leading
to a transport scenario as depicted in figure~\subref{resribcleangap}{c}: At these high disorder strengths the disordered region becomes more and more (Anderson) insulating,
and eventually acts almost like a perfect insulator. For an interface with a perfect insulator, however, one again has one-dimensional edge transport
of the (now relocated) topological edge states. As we discussed previously, those are not very susceptible to the effects of a magnetic field because
of their purely one-dimensional nature.

\begin{figure}[]
\centering
\includegraphics[width=0.75\textwidth]{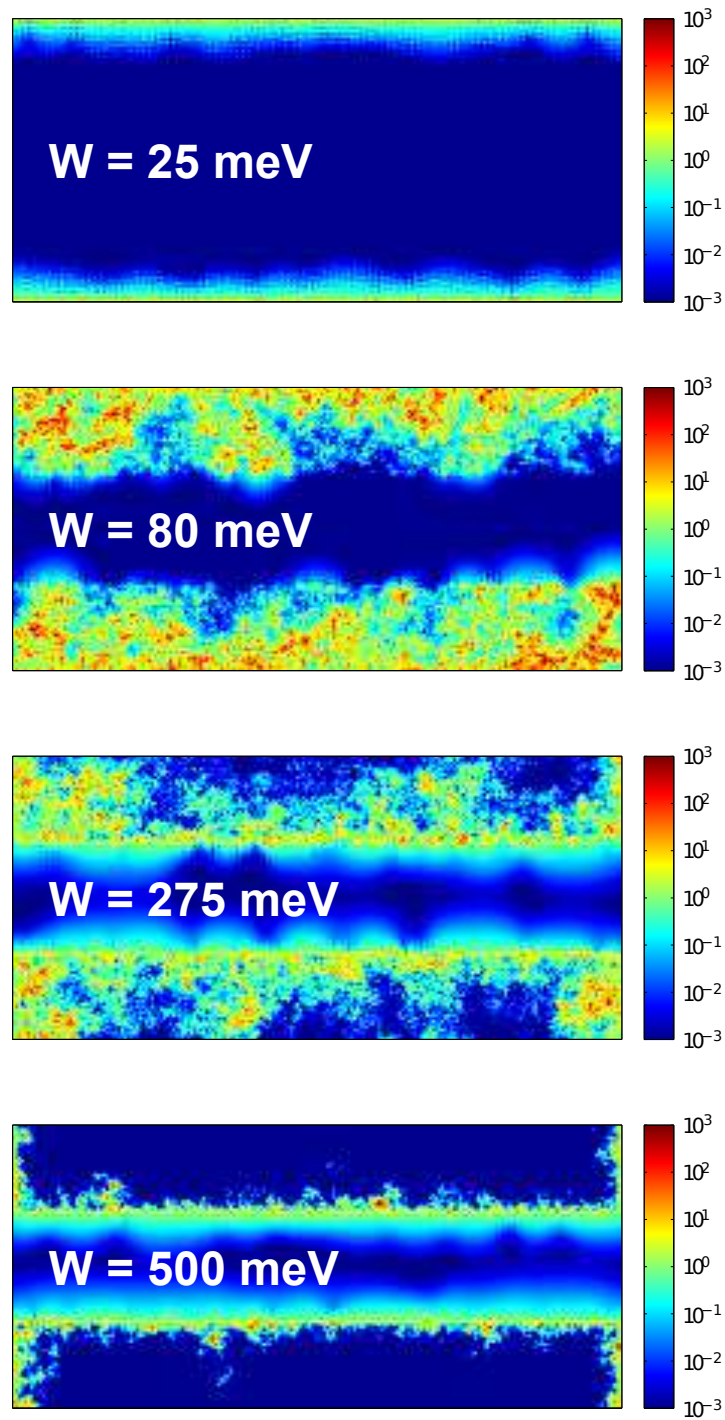}
\caption{ Color plots of the local density of states (arbitrary units) of the scattering region in a logarithmic scale for the setup shown in figure~\subref{resribcleangap}{a}
for different disorder strengths $W$. The results were obtained on a single disorder realization. They nicely show the transition from pure edge transport at the
outer boundary to edge transport along the disorder region (as illustrated in figure~\subref{resribcleangap}{c}) with the intermediate metallic regime in which
the current is carried all over the disordered region.
}
\label{resdisorderchannel}
\end{figure}

To see that we indeed obtain a scenario as depicted in figure~\subref{resribcleangap}{c}, one may look at snapshots of the local density of states for
a fixed disorder configuration, as shown in figure~\ref{resdisorderchannel}. Here, one clearly can see the described transition from pure edge transport
along the outer edge to another edge transport situation, where the strong disorder just acts as an effective Anderson insulator. Also one clearly observes
the intermediate metallic regime, where the transport is full delocalized over the whole disorder region (second panel in figure~\ref{resdisorderchannel}).

The fact that strong disorder regions effectively act like a trivial insulator with new edge states connected to the interface of these disorder regions
to clean topological insulators has already been observed in the context of three-dimensional topological insulators \cite{Schubert2012}. However, 
to our knowledge it has not been seen in a transport setup and it may show an alternative way of patterning such materials. E.g., one could think
of using ion beams to locally create strong disorder patches instead of etching away the material which should lead to similar results in terms
of edge state transport.

%
%
%
%

Summing up the results obtained so far, we find that the one-dimensional edge transport along the edges of a 2d topological
insulator turns out to be very robust against the combined effect of disorder and magnetic field. This is in line with
experimental results on smaller samples but differs from the numerical findings of \cite{Maciejko2010} in which
a strong response of the transmission was observed already at moderate disorder strengths and small magnetic fields. In this
publication they observe an almost linear dependence of the transmission on the magnetic field. This seems surprising at first
glance, as the calculations performed in \cite{Maciejko2010} were done with a similar method that is also used in
the present publication.

The main difference between \cite{Maciejko2010} and the study at hand is that \cite{Maciejko2010} makes
a small but seemingly important modification to the ribbon calculation setup that is shown in figure~\subref{geostrip}{a}, by
introducing an artificial mass term in the Hamiltonian for the uppermost row of cells used with the intention to additionally decouple
the edges similar to the clean gap that we used. They also use a very small clean gap of $30\,\mathrm{nm}$. We suspect that especially the mass term is responsible for the strong magnetic field
signature that they observe even at low disorder strengths. In our calculation, we also find this strong signature, but only in the parameter range where the disorder
strength is already large enough to lift the insulating behavior of the bulk and create an extended metallic region, i.e., in the range of $W>40\,\mathrm{meV}$
in figures~\subref{resrib}{a} and \subref{resribcleangap}{b}.

As already noted in \cite{Maciejko2010}, this strong almost linear response to the magnetic field seems to
agree with experimental results on larger samples. In view of our results, this suggests the existence of metallic patches in these long samples
which would lead to such a magnetoresistance signature. However, we do not find
it very likely that these metallic patches are due to strong inhomogeneous disorder, but they might come about due
to locally trapped charges which act like a local gate on the sample and in this way create small metallic puddles. The
existence of such puddles was also hypothesized by experimentalists and considered in other theories as they might be
responsible for the backscattering of the edge states at zero field which would explain the non-quantized transport in these samples \cite{Roth2009,Konig2013,Grabecki2013,Vayrynen2013,Vayrynen2014}.
As this is symmetry forbidden in the coherent single-particle picture, these theories focus on inelastic two-particle backscattering
or include additional processes which break the phase coherence of the wave function evolution.

In the following second part, we will study in more detail the influence of such metallic puddles in the framework of single-particle
scattering and focus on the combined effect of disorder, spin-orbit interaction and magnetic field in such a setup.

\subsection{Effect of local gating}

%
%
\begin{figure}[]
\centering
\includegraphics[width=\textwidth]{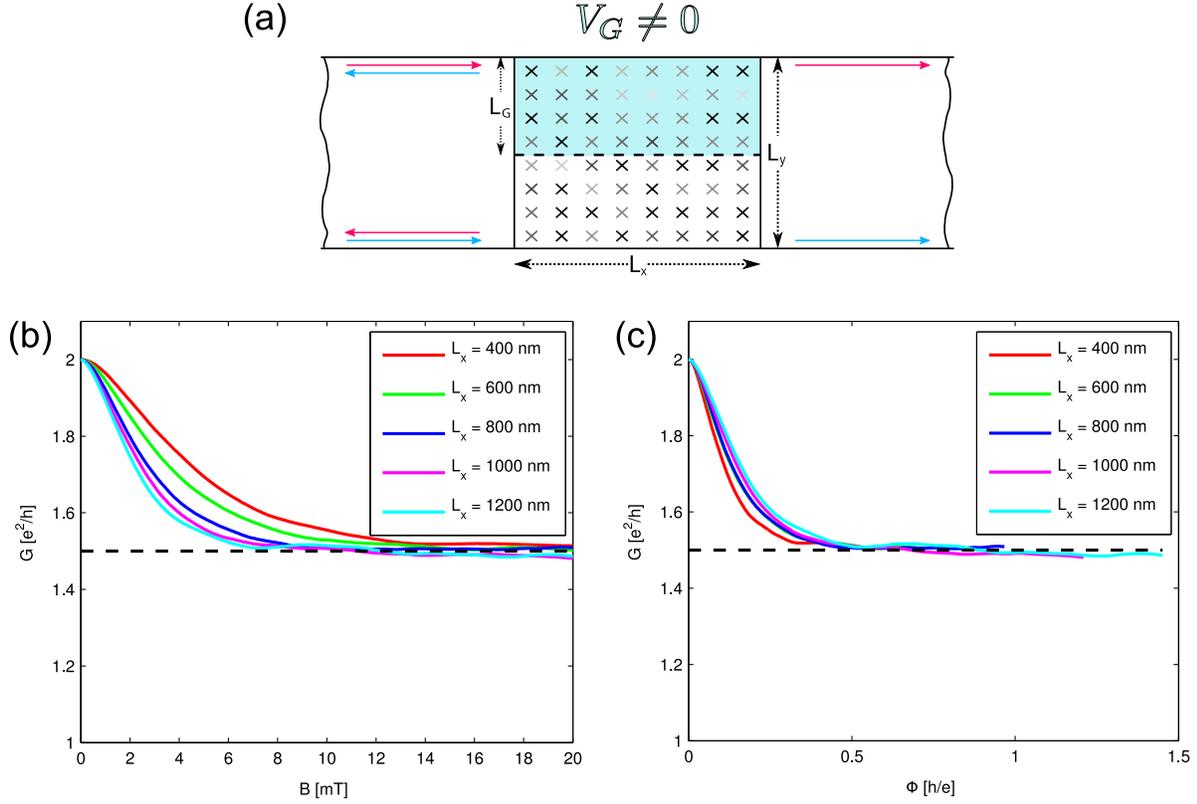}
\caption{ (a) Scheme of the ribbon geometry with an additional local gate that creates a metallic patch
on one edge of the sample. The other edge is unaffected and the two edges are decoupled by the insulating bulk, given that
the disorder is not chosen too strong. (b) Disorder averaged results for the transmission for the setup shown in (a) as 
a function of the magnetic field and different lengths $L_\mathrm{x}$ of the gated region, with width $L_G = 250\,\mathrm{nm}$, $L_y = 500\,\mathrm{nm}$
at a disorder strength of $W = 5\,\mathrm{meV}$ and a potential shift due to the gate of $V_G=50\,\mathrm{meV}$. The average was taken over 1000 disorder realizations.
(c) The same dataset as in (b) plotted as a function of the magnetic flux through the gated area, which shows that already
one flux quantum through the gated area is enough to trigger a transition to completely chaotic scattering in the puddle.
}
\label{resgatedifferentlength}
\end{figure}
%
%

To study the effect that a local metallic puddle would have in terms of the magnetoconductance in connection with
a small amount of disorder, we devise a calculation geometry that is shown in figure~\subref{resgatedifferentlength}{a}: A local
gate is added to one side of the ribbon, which creates a metallic puddle of length $L_x$ and width $L_G$. The lower edge
is not gated and its edge state propagates unaffectedly, as we will choose the disorder smaller than the critical
disorder needed to trigger the insulator-to-metal phase transition discussed above. 

Results for the transmission obtained
for such a geometry are shown in figure~\subref{resgatedifferentlength}{b} as a function of the magnetic field
for different lengths of the gated region. The disorder strength is chosen at a small value of $W = 5\,\mathrm{meV}$ with
a potential shift due to the gate voltage of $V_G=50\,\mathrm{meV}$. The drop in transmission with increasing $B$, in
presence of spin-orbit coupling $\Delta$, results from scattering paths as sketched in the bottom panel of figure~\subref{geostrip}{b}.
The results already resemble the ones obtained in \cite{Maciejko2010} and also the experimental results on
larger samples in the fact that they show a strong and almost linear behavior for small fields\footnote{The behavior very
close to $B=0$ is quadratic which agrees with other models for edge state localization \cite{Delplace2012} but it turns over to a linear dependence very quickly.}.
If one set
the gate voltage to zero, i.e., one were looking at true 1d edge states, one would not see any
response to the magnetic field at this disorder strength (cf. figure~\subref{resrib}{a}). This also explains why
the conductance obtained in our calculations is always larger than one as there is always the contribution
of the (almost) freely propagating edge state of the lower edge.
Looking more closely at the results, we see that the obtained curves all show a very similar behavior. In particular the
value for large fields is roughly $G\left(B\gg 1\right) \approx 1.5\,\textfrac{e^2}{h}$ for all calculated setups even though the
gated sections strongly differ in their length. This means that the edge states on one edge get reflected with a $50 \%$ probability
suggesting that the gated region shows the behavior of a chaotic cavity randomizing the exit direction. And indeed, in a closer analysis of the scattering
matrices connected to the transmission values we found that the submatrix of the scattering matrix that describes
the scattering of the upper edge states in this parameter range behaves as if it was drawn from the two-dimensional
\emph{circular unitary ensemble} CUE(2). In particular, this means that the transmission values obtained for
a single sample follow a box distribution on the interval $\left[ 0,1\right]$, i.e., they are completely random, which
on average leads to this $50 \%$ transmission probability.

Going from zero magnetic field to a finite field, we observe a transition from an on-edge scattering matrix which is
perfectly transmitting to a CUE(2)-matrix. The magnetic field scale on which this transition happens is given by
the number of flux quanta penetrating the gated region. This is best illustrated in figure~\subref{resgatedifferentlength}{c},
where the data from figure~\subref{resgatedifferentlength}{b} is replotted over the flux in the gated region measured
in units of $\textfrac{h}{e}$. One observes that now the data sets almost reproduce a single curve and show that already
a magnetic flux on the order of a single flux quantum is sufficient to obtain the chaotic scattering regime.

%
%
\begin{figure}[]
\centering
\includegraphics[width=\textwidth]{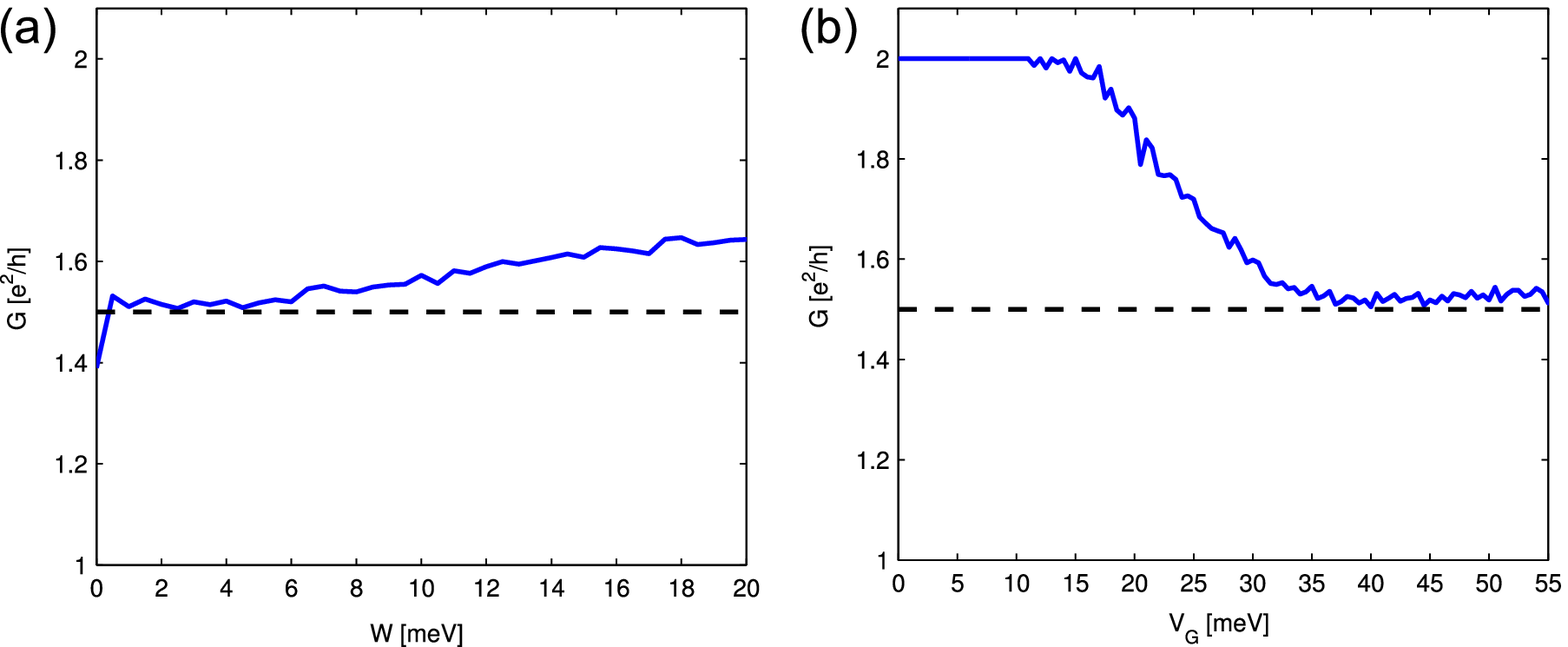}
\caption{ Disorder averaged results for the transmission of the setup with a one-sided gate 
(cf. figure~\subref{resgatedifferentlength}{a}) at a finite magnetic field, $B = 15\,\mathrm{mT}$, as a function of (a) the disorder
strength at a fixed gate voltage $V_G = 50\,\mathrm{meV}$ and (b) as a function of the gate voltage at fixed disorder strength
$W = 5\,\mathrm{meV}$. The lengths defining the device geometry are $L_G = 250\,\mathrm{nm}$, $L_y = 500\,\mathrm{nm}$ and $L_x = 1000\,\mathrm{nm}$.
One sees that the discussed chaotic scattering regime, $G \approx 1.5\,\textfrac{e^2}{h}$, is not just obtained for a single parameter set but can
be observed over a larger parameter range.
}
\label{resgatedifferentdisorderorgate}
\end{figure}
%
%

To show that this CUE-regime is obtained over a wider parameter range, we made additional calculations
in which we varied the disorder strength and the gate voltage, leaving the magnetic flux big enough to be in
the chaotic transport regime. Looking at the results which are shown in figure~\subref{resgatedifferentdisorderorgate}{a} and \subref{resgatedifferentdisorderorgate}{b}
one sees that one obtains an extended plateau exhibiting the characteristic CUE-transmission of $G = 1.5\,\textfrac{e^2}{h}$ showing
that this is indeed a quite generic case for low disorder strength and the puddle sizes considered here.

Given that the transport through a puddle has a wide parameter regime which shows this generic transport behavior it is tempting to
try to quantitatively relate this to the measured magnetoconductance of HgTe/CdTe quantum well samples. However, for this
one has to face the major complication that the experimental samples likely contain more than one single puddle
that we have considered so far. As shown in \cite{Delplace2012}, the logarithm of the transmission would be self-averaging
in such a series of scatterers and the averages of the logarithmic transmissions of single puddles would add up to yield the total logarithmic transmission, i.e.,
\[
\left\langle \ln T_{\mathrm{tot}}\right\rangle =\sum_{n=1}^N\left\langle \ln t_n\right\rangle.
\]
Assuming that the puddles are all of the same type and that we are restricting ourselves to the low
field behavior, in which the disorder averaged transmission of one puddle is close to one, it is roughly given by a linear
flux dependence for small fields,
\[
\left\langle t(\phi)\right\rangle=1-\frac{\phi}{\phi_c}, \label{eq:translinear}
\]
cf. figure~\subref{resgatedifferentlength}{c}, with the slope $1/\phi_c$.
Supposing that the sample has $N$ puddles on the edge and all have the same area $A$, 
one can infer the total edge transmission from the average transmission for a single puddle $\left\langle t(\phi)\right\rangle$:
\begin{eqnarray}
\exp\left\langle \ln T_{\mathrm{tot}}\right\rangle & = & \exp \left( N \left\langle \ln t(\phi)\right\rangle \right) \nonumber \\
 & \approx & \exp \left(-\frac{N\phi}{\phi_{c}}\right)\nonumber \\
 & = & 1-\frac{N\phi}{\phi_{c}}+\mathcal{O}(\phi^{2})\nonumber \\
 & = & 1-\frac{NBA}{\phi_{c}}+\mathcal{O}(B^{2}).
\end{eqnarray}
Thus, the slope of the conductance at low fields would be related to the product of the number of puddles and the typical puddle
area, i.e., the total area of all puddles. The experimentally extracted values for the slope of the conductance at small magnetic fields $\rmd [\sigma(B)/\sigma(0)]/\rmd B$ in long samples range from 
$\approx 0.5\,(\textfrac{1}{\mathrm{T}})$ \cite{Gusev2013} to $\approx 50\,(\textfrac{1}{\mathrm{T}})$ \cite{Konig2007}. 
If the puddles were in the universal CUE-regime, their transmission slope at zero field would be roughly given by $\phi_c\approx 0.5\,\textfrac{h}{e}$, cf. figure~\subref{resgatedifferentlength}{c}.
This would correspond to a total puddle area in the sample of $1000\,\mathrm{nm}^2$ resp. $100000\,\mathrm{nm}^2$, i.e., a total linear puddle size of only
$30\,\mathrm{nm}$ resp. $300\,\mathrm{nm}$ which seems to contradict the assumption of the puddles being in the  CUE-regime, as
this area would be too small for that\footnote{In our numerical calculations, we only observe the CUE-regime
at linear puddle sizes larger than $\approx 100\,\mathrm{nm}$ for $V_G=50\,\mathrm{meV}$. For smaller puddles there is a non-vanishing
chance that the electrons directly traverse the puddle without being scattered chaotically leading to an overall average transmission
which remains above $0.5$ even in the regime of ergodic dynamics.} if it would be not just one puddle.
So from the perspective of fully elastic coherent transport one could draw the conclusion that the puddles in the experimental samples should either
be shallower than $V_G=35\,\mathrm{meV}$, cf. figure~\subref{resgatedifferentdisorderorgate}{b}, or smaller than $100\,\mathrm{nm}$. Then their magnetoconductance
signature would still be similar to figure~\subref{resgatedifferentlength}{c} but it would saturate at a value above $0.5\,\mathrm{e^2}/{h}$, the
slope at zero field would be decreased and would be compatible with the experimentally determined slopes.

However, this argumentation breaks down if the puddles present in the sample already lead to backscattering even at zero field, which is one hypothesis to explain
the non-quantized zero field resistance in long samples. This could be due to the transport being partly incoherent \cite{Roth2009} or due to Coulomb interactions \cite{Vayrynen2013,Vayrynen2014}, both
of which are not included in this work but would --- given the processes are sufficiently strong --- also lead to an asymptotic 50-50 backscattering probability of the puddle. 
The additional effect of a magnetic field acting on such a puddle which already at zero field starts out closer to this 50-50 limit is expected to be strongly reduced, i.e., $\phi_c>0.5\,\textfrac{h}{e}$  which
would also allow the existence of larger puddles in such samples. We are currently preparing a quantitative study on the influence of dephasing on the zero field
backscattering \cite{newmanu}.

As seen on the above argumentation, it is still difficult to make definite statements about the situation in current experimental samples. We therefore see
a need for more experiments on small well defined systems in which one studies the influence of a single artificial puddle by adding a local gate.
With this, the applicability of the presented coherent elastic theory could be verified which would provide a solid basis for the interpretation
of results in longer samples.

\section{Conclusion}
Using a model which includes only elastic coherent disorder scattering, we find qualitative agreement with magnetoconductance
results on long samples of HgTe/CdTe quantum wells (acting as a 2d TI) if we assume the existence of charge puddles in the material. Without
charge puddles one finds the edge states to be uninfluenced by the combined effect of magnetic field, spin-orbit coupling and disorder. This changes only
if one increases the disorder to a strength at which the material is already on the phase transition to a topologically trivial Anderson insulator,
a scenario which we do not find very likely to occur in the available materials as it would imply that one should not be able to observe
the measured quantized conductance and non-local transport phenomena.
This is in line with other numerical studies which also obtained robust edge states under the combined influence of magnetic field and
disorder \cite{Pikulin2014,Chen2012a} but disagrees with \cite{Maciejko2010} in which a magnetoconductance signature similar
to our results for a charge puddle is observed in a setup without puddles, i.e., only considering magnetic field and disorder. However,
this publication uses a modified calculation setup including an artificial mass term which might modify the magnetoconductance
signature in a way that they are probing the regime of very strong disorder.

The interpretation that the puddles are responsible for the measured magnetoconductance signature would also be in line with the observation
that the quantized conductance observed in short samples is robust under the influence of an applied field. One could argue that
there is a chance that the short samples which show the quantized conductance are small and clean enough not to contain any
charge puddles while this would be statistically unlikely for larger samples.

The presented theory is still incomplete in the sense that it can not account for the observed backscattering (non-quantized conductance)
at zero field that is ubiquitous in all measurements on long HgTe/CdTe quantum well samples. However, a number of theories also
ascribes this backscattering to the existence of puddles due to their inherent dephasing \cite{Roth2009} or their enhanced
electron-electron backscattering \cite{Vayrynen2013,Vayrynen2014}. These effects (backscattering, loss of electron phase coherence), however,
are not expected to be strongly influenced by the magnetic field. It might therefore well be that this elastic coherent theory accounts
for the influence of the magnetic field while the zero field backscattering is due to other processes, the common root of both being
the existence of charge puddles.
Without further experiments however, this remains highly speculative. We therefore strongly recommend to focus the experimental effort
especially on short samples which show the quantized longitudinal conductance at zero field\footnote{Published results for the magnetoconductance of such a HgTe/CdTe sample would be desirable}.
In these samples one could introduce artificial puddles by adding a local gate with which one could check the predictions
of this manuscript and therefore the validity of the BHZ model and the applicability of a fully coherent elastic theory.
For such a test experiment, we recommend the discussed CUE-regime, a wide parameter regime in which the puddle is
showing ``maximal conductance fluctuations'', i.e., fully random transmission values $t\in [0,1]$. As this is expected for a wider range
of gate voltages and magnetic fields one should be able to detect this regime even in a single sample without the need for an
additional disorder average. In addition, one would also be able to check the influence of the puddle on the transmission
at zero field which could provide valuable clues about the backscattering mechanism in long samples\footnote{We are currently
preparing a quantitative study about the influence of dephasing in a single charge puddle \cite{newmanu}.}.

So far we have not done any calculations on the other very promising material system InAs/GaSb. However, we would expect
that it behaves similarly under the application of an external gate, i.e., we also expect the existence of a CUE
regime, and this has indeed already been seen in other numerical simulations \cite{Mi2013}.

In passing, we observed that the phase transition caused by very strong disorder might provide a new
way of patterning the samples: Instead of removing the unwanted material, one could consider to strongly disorder
it, e.g., by ion beam irradiation to render it topologically trivial and thereby modify the current path of
the topological edge states.

\ack
This work is supported by Deutsche Forschungsgemeinschaft
(SPP 1666 and joined DFG-JST Research Unit FOR 1483).
We thank V.~Krueckl for many helpful discussions throughout the course of this work and I.~Adagideli as well as M.~Wimmer for useful conversations.

\end{document}